 \documentclass[final,5p,times,twocolumn,authoryear]{elsarticle}

\usepackage{amssymb}
\usepackage{lipsum}
\usepackage{hyperref}
\usepackage{listings}

\newcommand{\og}{Open\textsc{Gadget3} }

\journal{Astronomy $\&$ Computing}

\begin{document}

\begin{frontmatter}



\title{\og GPU solver tests}


\author[inaf-ts]{A. Ragagnin}
\author[usm]{G. S. Karademir}
\author[usm]{F. Groth}
\author[usm,mpa]{K. Dolag}
\author[uchicago]{L. M. B\"oss}
\author[ifpu,iscs,inaf-ts,infn-ts]{T. Castro}
\author[intel]{N. Hariharan}
\author[difa]{M. Aiello}
\author[inaf-ts]{L. Tornatore}

\affiliation[inaf-ts]{organization={INAF-Osservatorio Astronomico di Trieste, Via G. B. Tiepolo 11, 34143 Trieste, Italy}}
\affiliation[usm]{organization={Universitäts-Sternwarte, Fakultät für Physik, Ludwig-Maximilians-Universität München},
            addressline={Scheinerstr. 1},
            city={81679 München},
            %postcode={},
            %state={},
            country={Germany}}
\affiliation[mpa]{organization={Max-Planck-Institut für Astrophysik, Karl-Schwarzschild-Straße 1, 85741 Garching, Germany}}
\affiliation[uchicago]{organization={Department of Astronomy and Astrophysics, The University of Chicago, William Eckhart Research Center, 5640 S. Ellis Ave. Chicago, IL 60637, USA}}
\affiliation[ifpu]{organization={IFPU - Institute for Fundamental Physics of the Universe, Via Beirut 2, 34014 Trieste, Italy}}
\affiliation[iscs]{organization={ICSC - Italian Research Center on High Performance Computing,
Big Data and Quantum Computing, Italy}}
\affiliation[infn-ts]{organization={INFN, Sezione di Trieste, Via Valerio 2, 34127 Trieste TS, Italy}}
\affiliation[intel]{organization={Intel Technology India Private Ltd, 23-56P, Devarabeesanahalli, Varthur Hobli, Outer Ring Road, Bangalore - 560103, KA, India}}
\affiliation[difa]{organization={Dipartimento di Fisica e Astronomia "Augusto Righi", Alma
Mater Studiorum Università di Bologna, via Gobetti 93/2, I-40129
Bologna, Italy}}

\begin{abstract}
We present an in-depth evaluation of the scalability and accuracy of the GPU porting of the N-body code for hydrodynamic cosmological simulations \og. While technical details of our GPU porting were presented in \cite{Ragagnin2020GPU}, in this work we focus on assessing the accuracy of the ported modules: the short range gravity integrator, the different components of the hydrodynamic solver, and the conjugate gradient solver for thermal conduction. We ran several tests that gradually increase the number of physical modules included: a gravity-only cosmological simulation; a hydrodynamical shock tube test; a non-radiative zoom-in simulation of a galaxy cluster in a cosmological box; and a full-physics zoom-in simulation of a galaxy in a cosmological box. Comparing the results obtained with the GPU implementation to those from the classical CPU version, we find excellent agreement across all tests, with small differences on very small scales. For the individual physical modules, we find a GPU chip-to-chip speedup ranging from $\approx3-5$. For more complex cosmological and hydrodynamical setups, where a large number of physical processes and overheads contribute to the total workload, the observed total chip-to-chip speedup (with the same number of nodes and CPUs per node) is $\approx2-3$. We ran our tests on four different supercomputers: Leonardo Booster (CINECA), MareNostrum-V (BSC), SuperMUC-NG2 (LRZ), and the CIP cluster of the Faculty of Physics at the Ludwig-Maximilians-Universit\"at (LMU).
\end{abstract}




\end{frontmatter}

%\tableofcontents



\section{Introduction}

\begin{table*}[b]
    \centering
    \begin{tabular}{r l l l l l}
    \hline
    Facility & Machine & Host & Memory & Device & Stack\\
    \hline
    CINECA & Booster & $1\times$ Intel Xeon $32$ cores & 512 GB & $4\times$ Nvidia custom Ampere $64$GB  & cuda 12.3\\
    &&&&&nvhpc 23.11\\
    &&&&&openmpi 4.1.6\\
    &&&&&gsl 2.7.1\\
    &&&&&fftw 3.3.10\\
    \hline
    LRZ & SuperMuc NG2 & $2\times$ Intel Sap. Rapids $52$ cores & 512 GB & $4\times$ Intel Ponte Vecchio $128$ GB & Intel MPI 2021.12\\
    &&&&&(gcc 7.5.0)\\ 
    &&&&&gsl 2.8\\
    &&&&&fftw 3.3.10\\
    \hline
    BSC & MareNostrum 5 & $2\times$ Intel Sap. Rapids $52$ cores & 512 GB & $4\times$ Nvidia Hopper GPU 64  GB & cuda 12.9\\
    &&&&&nvhpc 25.7\\
    &&&&&openmpi 4.1.6\\
    &&&&&gsl 2.7.1\\
    &&&&&fftw 3.3.10\\
    \hline
    LMU & CIP & $32\times$  Intel Xeon & 128 GB & $4\times$ Nvidia H100 GPU & cuda 12.4\\
    &&&&&nvhpc 24.3\\
    &&&&&openmpi 5.0.0\\
    &&&&&gsl 2.7.1\\
    &&&&&fftw 3.3.10\\
    \hline\\
    \end{tabular}
    \caption{Current configurations of the individual nodes of the two systems used for testing.}
    \label{tab:clusters}
\end{table*}

Contemporary high-performance computing (HPC) facilities are increasingly employing computing nodes equipped with accelerators, such as Graphics Processing Units (GPUs).
At the time of writing, nine out of the top ten leading Top500\footnote{\url{https://www.top500.org/}} HPC facilities have transitioned to GPU adoption. This shift underscores the critical importance of optimising existing scientific codes to exploit GPU-based architectures. Similar efforts are ongoing for various astrophysical codes, including 
AthenaK~\citep{Stone2024AthenaK}, 
Bonsai~\citep{Bedorf2012b},
CRK-HACC~\citep{Frontiere2023HACC},
DISCO-DJ~\citep{List2025}.
MP-Gadget~\citep{Feng2015MPGadget,Feng2018MPGadget},
GAMER~\citep{schive2018gamer}, 
gPLUTO~\citep{Mignone2012},
AREPO~\citep{Springel2010Arepo,Zier2024ArepoGPU}, 
SPH-EXA~\citep{Cavelan2020,Keller2023},
and
pkdgrav3~\citep{Meier2025Pkdgrav3GPU}.

This manuscript focuses on accuracy tests for the GPU porting of the N-body code for hydrodynamic cosmological simulations \og~\citep[used for instance in][]{Marra2022,
Marin-Gilabert2022,
Dolag2023,
Fischer2023,
Groth2023,
Boess2024,
Chaitra2026,
Damiano2026}\footnote{\url{https://www.space-coe.eu/codes/opengadget.php}}, which is the latest version of Gadget3, a successor of Gadget2~\citep{Springel2001Gadget, Springel2005Gadget2}. The public release of \og is one of the milestones within the European Centre of Excellence SPACE\footnote{\url{https://www.space-coe.eu/}} \citep[][]{SPACE2025} and is scheduled for the end of 2026.

Our implementation is capable of running several \og modules on the GPU: the short-range gravitational interaction~\citep{BarnesHut1986}; the hydrodynamic density and force computations using smoothed particle hydrodynamics~\citep[SPH,][]{Gingold1977SPH}; and thermal conduction. The porting can utilise either OpenACC\footnote{\url{https://www.openacc.org/}}~\citep[see e.g.][]{OpenACC} or OpenMP-target directives~\footnote{\url{https://www.openmp.org/spec-html/5.0/openmpsu60.html}}~\citep{OpenMP}, although the OpenMP-target implementation has so far only been tested for the Barnes-Hut interactions.

We performed our tests on different systems, namely CINECA Leonardo Booster\footnote{\url{https://leonardo-supercomputer.cineca.eu/it/leonardo-hpc-system/}}, the LRZ SuperMuc-NG2\footnote{\url{https://doku.lrz.de/supermuc-ng-10745965.html}}, BSC MareNostrum 5\footnote{\url{https://www.bsc.es/marenostrum/marenostrum-5}}, and the CIP cluster of the Faculty of Physics at the LMU \footnote{\url{https://www.physik.lmu.de/en/about-us/facilities/it-service/}}. The configuration of the individual nodes is listed in Table \ref{tab:clusters}.

The paper is organised as follows: in Section \ref{sec:infra}, we review the infrastructure and porting details of \og relevant to this paper. Section \ref{sec:tests} is dedicated to an in-depth study of various test cases, carefully selected to gauge the accuracy of the GPU-accelerated \og in hydrodynamic and cosmological simulations. In Section \ref{sec:progress} we present weak scaling test results. Finally, we draw our conclusions and discuss future prospects in light of code profiling in Section \ref{sec:conclu}.

\section{\og infrastructure}
\label{sec:infra}

\begin{figure}
\centering
\includegraphics[width=\linewidth]{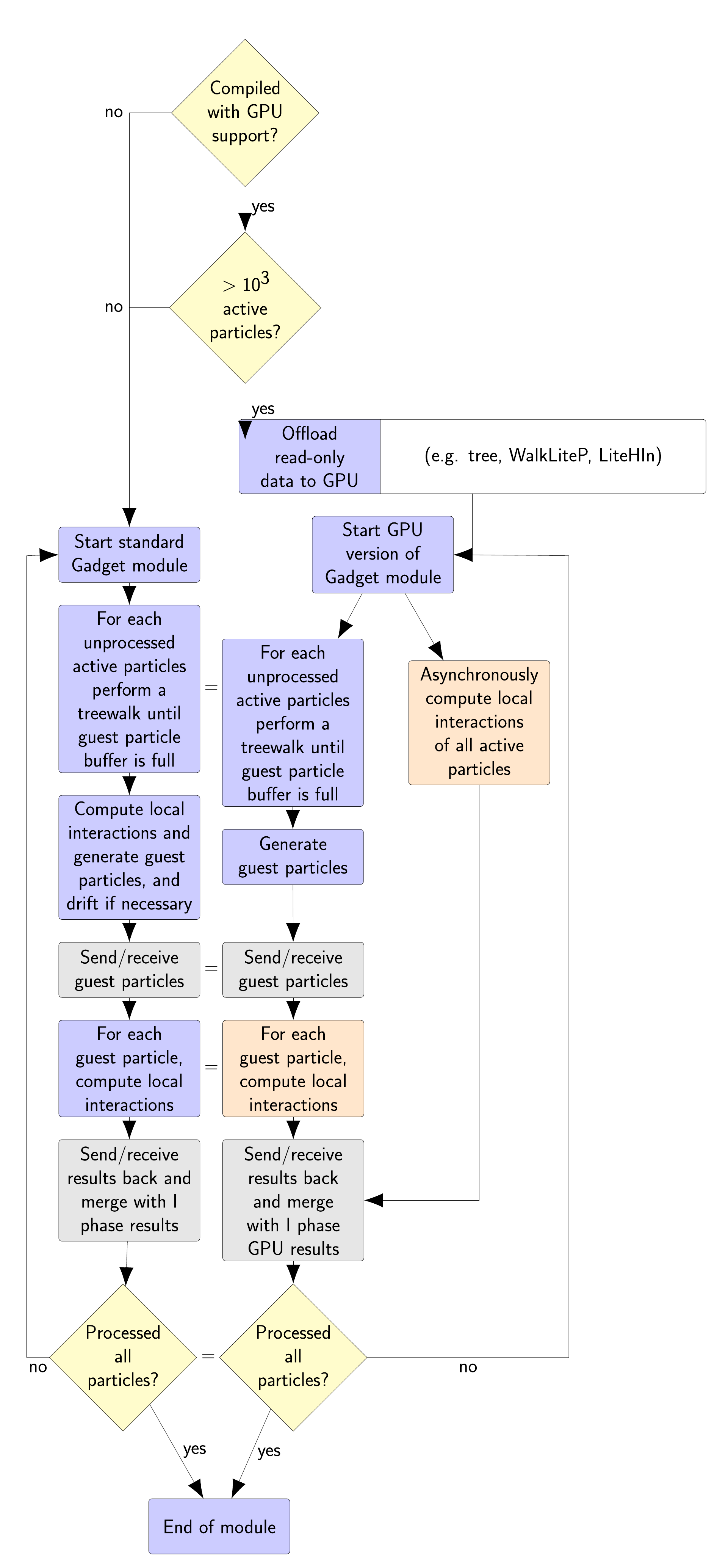}
\caption{\label{fig:flowchart}Flowchart summarising the GPU porting paradigm as implemented in~\cite{Ragagnin2020GPU}, for the particular case of the hydrodynamic force implemented in the file \texttt{Hydro/hydra.cpp}. Yellow rhombic shapes indicate decisional branches; blue rectangles briefly describe algorithms executed in the order following the direction of the arrows; orange rectangles indicate algorithms executed on the GPU; grey rectangles indicate algorithm blocks dominated by MPI calls. Note that the left part of the diagram schematically reports the hydrodynamic force computation module as performed by the standard \textsc{Gadget2/3} code, while the right part reports the implementation for GPU-active timesteps. To ease reading, left and right blocks with the same text are connected with an equal sign ``$=$``.}
\end{figure}

In this Section, we briefly outline the pertinent features of \og relevant to this manuscript.
While the porting strategy is described in detail in \cite{Ragagnin2020GPU}, we briefly outline the key details in the following to help readers understand the improvements achieved and to better evaluate the accuracy tests in the next Section.

The \og code inherits its code infrastructure from \textsc{Gadget3}~\citep{Springel2005Gadget2}.
 Over time, \textsc{Gadget3} was extended to have the capability to operate across a large number of CPUs, using a hybrid, MPI and OpenMP parallel scheme and allowed the code to scale up and use all computing resources of supercomputing systems~\citep{Hammer2016}.

Short-range interactions, such as the~\cite{BarnesHut1986} gravity solver, SPH density finder, SPH gradient computation, SPH hydrodynamical force computation, and thermal conduction computation, share the underlying structure where, for each dynamically active particle, a tree walk is performed to compute the forces acting on it.
Every module execution is divided into two phases: the primary phase computes same-rank interactions and produces an export buffer for particles that need to interact with domains on different ranks; while the secondary phase, after receiving an array of guest particles from other MPI ranks, computes guest particle interactions with local neighbours.
Due to total memory limitations, it may be necessary for the code to iterate through the primary and secondary phases multiple times to finalise processing of all active particles.

The SPH density module has an additional level of complexity as it operates as an iterative process, aiming to determine a smoothing length containing a fixed number of neighbours. Consequently, the aforementioned primary and secondary phases are iterated multiple times until convergence. Additionally, the density module can search for neighbours of star and black hole particles; their local density is used in the respective subgrid physics modules.

It is worth noting that there is a technical difference in the neighbour processing between the Barnes-Hut algorithm and the other algorithms (SPH density, SPH hydro force, and thermal conduction).
The former computes the impact of neighbours on a target particle for each isolated neighbour fetched by the tree walk and opened with an angular criterion; while each of the latter modules first collects all neighbours within a sphere around the target particle in an array, and processes them in a separate "evaluate" function.
This last detail allows for an additional performance saving by performing fewer tree walks by grouping nearby particles together, as implemented in~\cite{Ragagnin2016GreenTree}.

\subsection{\og GPU porting}
\label{sec:porting}

In Figure~\ref{fig:flowchart}, we schematically report the main points of each code module and the porting strategy. In particular, for simplicity, we limit our scheme to the porting of the hydrodynamic force (and we highlight the name of the corresponding file, \texttt{Hydra/hydra.cpp} for completeness); we refer to~\ref{ap:code} for more technical details on the actual code used for the porting.

As can be seen from the top part of Figure~\ref{fig:flowchart}, in our GPU porting, we divide the primary phase into two asynchronous tasks: while the CPU assembles the list of target particles and communicating tree nodes, the GPU executes the local interactions.
Subsequently, while the GPU is still processing local interactions (namely, within the same MPI rank), the CPUs exchange guest particles.
In the secondary phase, the guest particles can be handled by the GPU if it has already completed its task (from the primary phase) and received enough particles (otherwise, they are processed by the CPU). It is noteworthy that in our prior work~\citep{Ragagnin2020GPU}, we identified suboptimal GPU performance in time bins with only a few hundred active particles. Consequently, timesteps containing fewer than $1000$ particles are completely executed on the CPU (we stress that this is equivalent to running the non-ported, standard code), as depicted in the left branch in Figure~\ref{fig:flowchart}, while timesteps running on the GPU follow the right branch of the scheme reported in Figure~\ref{fig:flowchart}.

The CPU scheme for the drift of particles is adaptive, where active particles are drifted at the beginning of the timestep, while the others are drifted when encountered during a tree walk.
The latter type of drift requires a tread-locking operation (namely, to use \texttt{\#pragma omp critical} OpenMP regions) in order to avoid a data race during the position drift.
Since this approach would be problematic on GPU,  we execute the particle drifting process at the onset of a GPU-ported timestep for all particles (i.e., active and non-active ones).
While this adds a small computational overhead (negligible for timesteps with more than 1000 particles), it also eliminates several OpenMP critical regions in the interaction between active and non-active particles.
This should thus improve performance on GPUs, since they are based on the single-instruction multiple threads (SIMT) paradigm. However, this inevitably adds small discrepancies between the CPU and GPU versions of the code.

Moreover, to best exploit SIMT GPU parallelisation, we changed the way neighbours are processed: in the standard CPU version, the code first finds all local neighbours for a given particle and then computes their impact on the target particle. In our porting, we find neighbours in chunks of $32$ neighbours and process their impact on the target particle for these $32$ neighbours only before searching for a new chunk of neighbours.
We were able to implement this change by exploiting the fact that Gadget-like codes already have infrastructure capable of processing multiple chunks of neighbours, since it has to combine guest-particle results coming from neighbouring MPI ranks.

Gadget-like code kernels typically share the following paradigm: a list of active particles interacts with neighbours, of which we typically need to read only a subset of properties (e.g., position, temperature, etc.); the active particles, on the other hand, besides a subset of properties that we need to read (which are the same as for neighbours), also have a set of properties that have to be computed and stored (e.g., force acceleration).
Since our porting adopts the same data structures as the Gadget secondary phase, we are able to minimise memory transfer by sending only properties that need to be read and downloading only properties that have to be stored.
Moreover, we offload the tree at each timestep (remember that we drift all particles at the beginning of the timestep). The same holds for particle properties such as masses, positions, and velocities, which are updated only at the end of the timestep.

While in the first version of the porting presented in~\cite{Ragagnin2020GPU} we used only OpenACC, we now support OpenMP target constructs. Most \texttt{\#pragma} conversions between the two standards are straightforward, but they differ in how they handle asynchronous computation: OpenACC provides built-in functionality to check if a GPU asynchronous computation has completed, \texttt{acc\_async\_test(...)}. This check is important, as it determines whether we offload particles received in the secondary phase to the GPU or CPU, depending on whether the GPU is still busy performing the asynchronous computation of the primary phase. The OpenMP-target standard currently does not provide a built-in equivalent of \texttt{acc\_async\_test(...)}. While one could implement this functionality by explicitly crafting additional OpenMP-task \texttt{depend}\footnote{\url{https://www.openmp.org/spec-html/5.0/openmpsu99.html}} clauses, at the time of writing, since it is in development, the OpenMP-target test ran with only the primary phase on GPU.

\section{Tests}
\label{sec:tests}

\begin{figure}
{\includegraphics[width=1.05\linewidth]{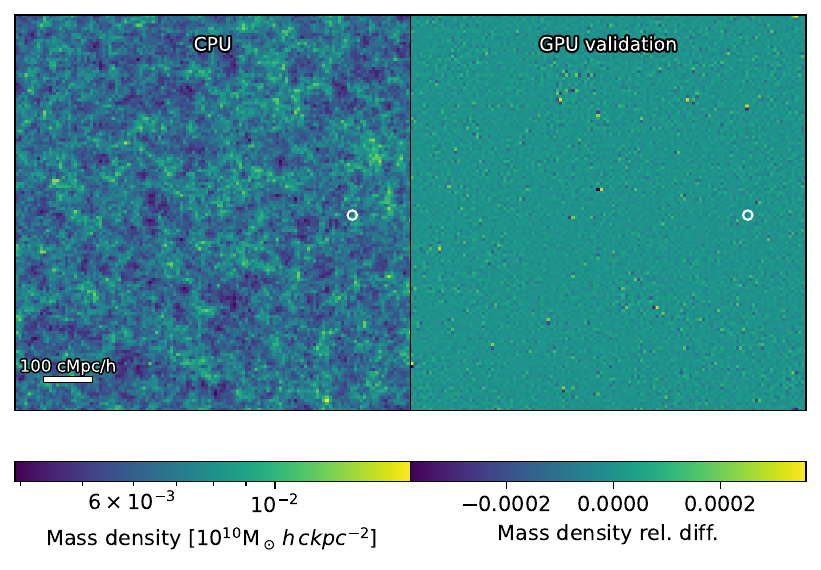}\\
\includegraphics[width=1.05\linewidth]{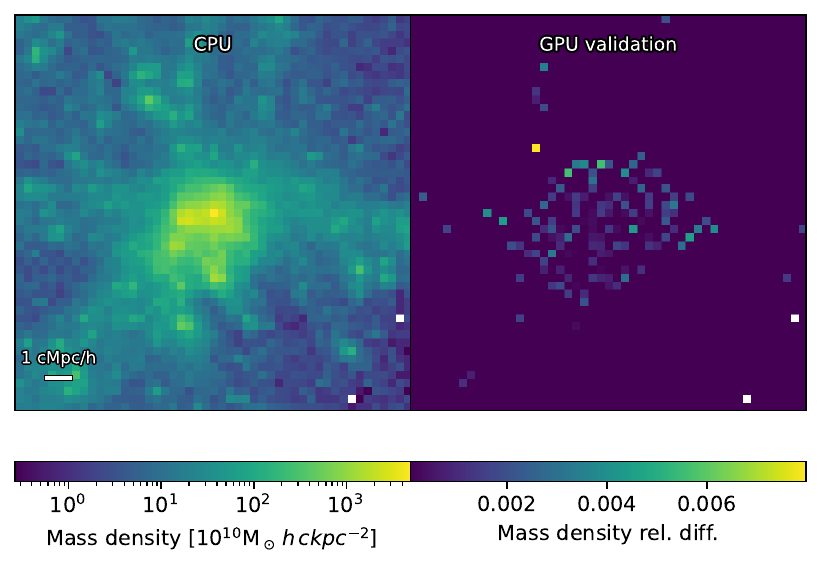}\\
\includegraphics[width=1.05\linewidth]{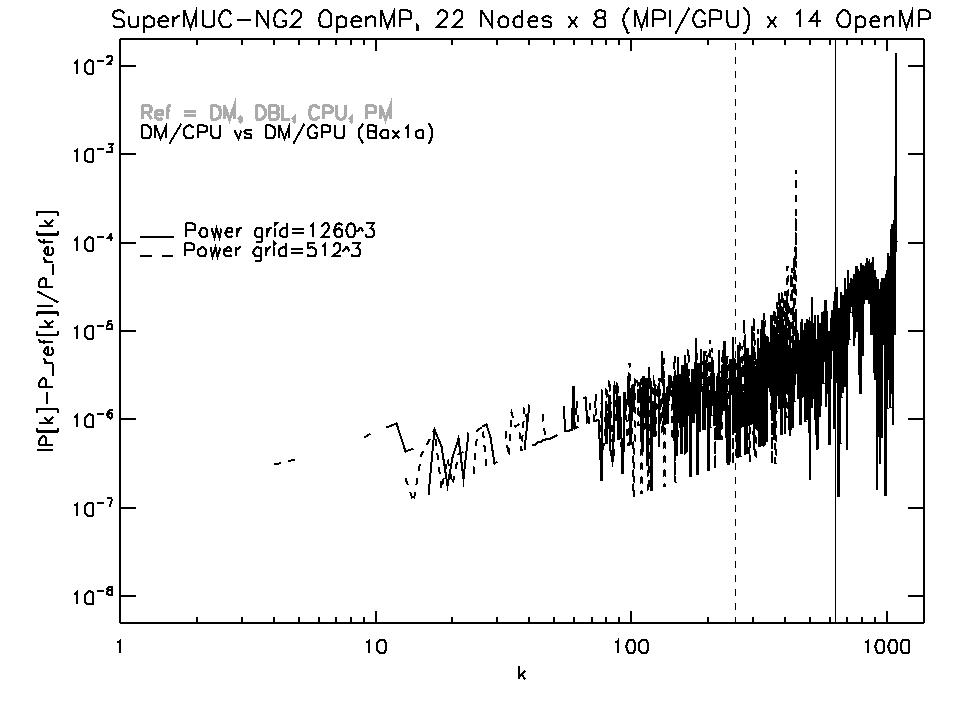}
}
\caption{\label{fig:dmo_scatter}
Comparison between CPU and GPU runs of the DMO cosmological box {\em Magneticum} {\em Box1a/mr}.
Top panels: Projected mass map of the complete box for the CPU (left) and GPU (right) run, with the most massive cluster highlighted with a white circle.
Central panels: Projected mass map of a $20\,{\rm cMpc}$ region around the most massive halo of the box. Bottom panel: Percentage difference between GPU and CPU power spectrum.}
\end{figure}

In this Section, we assess our GPU porting of \og through a dark matter-only simulation, focusing on gravity and tree-walking control on the GPU (using the OpenMP-target porting). Subsequently, we run three hydrodynamic tests adapted from \cite{Groth2023}.\footnote{\cite{Groth2023} provide initial conditions, configuration files, and plotting routines here:  \url{https://github.com/fgroth/hydro_tests}}
We then test both gravity and hydrodynamics (using OpenACC porting) by running a shock tube, and we test thermal conduction by running a small galaxy cluster simulation, both with non-radiative physics and one involving star formation, stellar feedback, and supermassive black hole growth and evolution.
See Table~\ref{tab:clusters} for a summary of the HPC architecture and library stacks used on each machine.

\subsection{Dark matter only run}
\label{sec:dmo}

To test the~\cite{BarnesHut1986} gravity integrator\footnote{In this setup, we test it in combination with the long-range gravitational forces from the PM, which uses \texttt{FFTW}~\citep[we used version 3.3.10, see][]{FFTW05}, that runs on CPU only.
Moreover, the time stepping integrator uses the external library gsl (we used version 2.7), see: \url{https://www.gnu.org/software/gsl}} we used a dark matter-only (DMO) initial condition from Magneticum\footnote{\url{www.magneticum.org}} Box1a/mr.
The simulation is initialised with $2\times1512^3$ particles within a volume of $869$ comoving Mpc (hereafter, cMpc), featuring dark matter particle masses of $m_{\rm DM}=1.3\times10^{10}\,h^{-1}\,{\rm M}_\odot.$
We also use the Gadget FoF/SubFind structure finder~\citep{Davis1985FoF,Springel2001Gadget,Dolag2009}, and the results comprise approximately $2.2\times10^4$ well-resolved halos (with more than $10^4$ particles per FoF group).

We ran this simulation on $22$ nodes on SuperMuc-NG2, where the CPU run used 8 MPI ranks per node, each with 6 OpenMP threads, while the GPU run used one additional GPU tile per MPI rank.
Note that this is the only test case where we used the OpenMP-target version of the porting to exploit SuperMucNG-2 accelerators.

The time-to-solution of the CPU was $10.2\times10^{4}\,{\rm s},$ and it spent $7.0\times10^{4}\,{\rm s}$ on the Barnes-Hut computation (see first row in Table~\ref{tab:tests}).
In comparison, the GPU run took $4.3\times10^{4}\,{\rm s}$ (speedup of $\sim2.3$), and it spent $1.7\times10^{4}\,{\rm s}$ (speedup of $\sim4.1$) on the Barnes-Hut computation.

Figure \ref{fig:dmo_scatter} top panel displays a projected density map covering the complete volume of the simulation, where we highlight (with a white circle) the position of the most massive object in the simulation. There, we can qualitatively appreciate that the CPU (left panel) and GPU (right panel) runs produce the same results at large scales, with errors in the mass content of $\approx0.02\%.$
In the central row of Figure \ref{fig:dmo_scatter}, we focus on a sphere of radius 1 cMpc, centred around the most massive halo in the simulation. Notably, both the central structure and the surrounding smaller haloes maintain consistent positions, with few pixels having non-zero (yet under one percent) differences, as expected from the slightly different time-stepping scheme.

To quantify these differences, we examine the power spectrum $P(k)$, defined as the auto-correlation function from the Fourier Transform $\delta({\bf k})$ of the density contrast $\delta({\bf r}) = \rho({\bf r})/\rho_c - 1.$ Here, $\rho({\bf r})$ represents the matter density at the coordinate  ${\bf r}$, and $\rho_c$ is the critical density of the Universe.
Figure \ref{fig:dmo_scatter} (bottom panel) illustrates our findings, presenting the total matter power spectrum $P(k)$ of the GPU run relative to that of the CPU run. The results confirm excellent agreement between the two runs, with sub-percent agreement extending down to the resolution of our power spectrum, approximately $200\,{\rm ckpc}$, derived from a grid of $512$ and $1260$ cells per side. The power spectrum differences due to changing grid size are of the same order of magnitude as the difference between CPU and GPU runs, showing that these differences are negligible. More precisely, they are of the order of $<0.01\%.$

\subsection{Ryu Jones shock tube}
\label{sec:sod}

\begin{figure}
\includegraphics[width=1.05\linewidth]{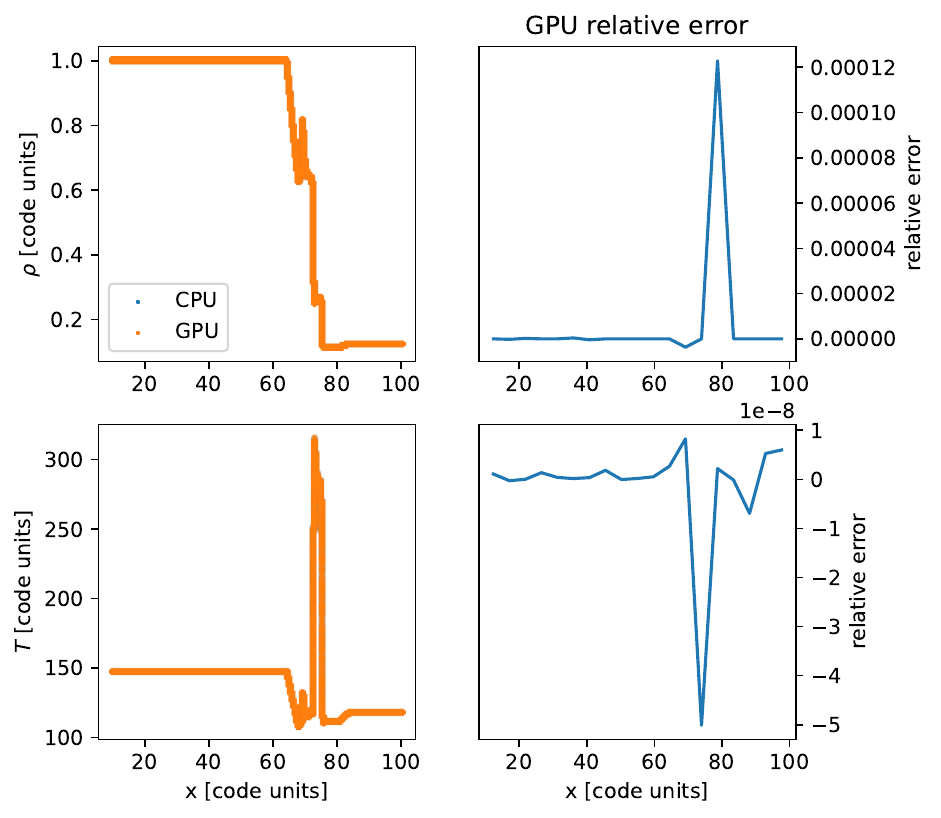}
\caption{\label{fig:shock_tube}Comparison of shock tube solutions. Each row reports the following quantities in code units: density $\rho$, temperature $T.$
}
\end{figure}

We test the accuracy of the GPU porting of the hydrodynamic solver by assessing its accuracy in resolving a \cite{RyuJones1995} magneto-hydrodynamic (MHD) shock tube.
This test therefore makes use of the SPH+MHD implementation~\citep{Wendland1995,Dolag1999,DolagStasyszyn2009,Dehnen2012,Pakmor2012,Price2012,Donnert2013,Beck2016,Tricco2016} of \og.

We ran a test case with $\approx2.1\times10^4$ particles on 1 node of the LMU CIP cluster, utilising one MPI rank with 24 OpenMP threads (see second row in Table~\ref{tab:tests}). The time spent by the CPU run was $4.4\times10^{3}\,{\rm s}$ (most of it spent in the SPH computations), while the GPU run (using 1 GPU per MPI rank) took $1.7\times10^{3}\,{\rm s},$ for a total speedup of $2.6.$
Figure \ref{fig:shock_tube} shows the final thermodynamic properties of the test case for both CPU and GPU. Here, we can see that the two codes produce extremely similar results, with a relative deviation in the density that is generally less than $0.001\%,$ and reaches $0.01\%$ only near discontinuities.

\subsection{Non-radiative galaxy cluster}
\label{sec:norad}

\begin{figure}
\centering
\includegraphics[width=0.8\linewidth]{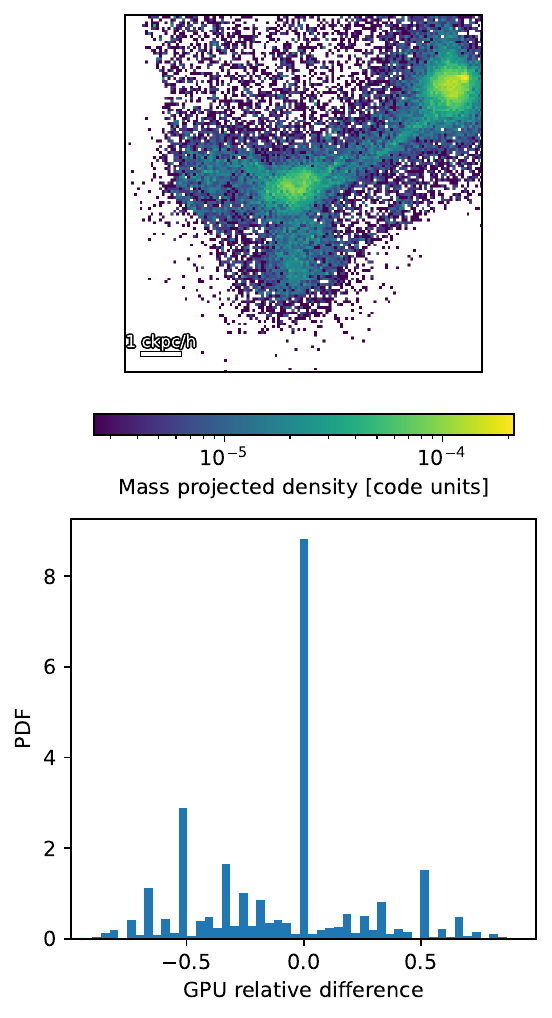}
\caption{\label{fig:norad_map}Non-radiative galaxy simulation density 2D map (top panel), and GPU relative difference (bottom panel).
}
\end{figure}

\begin{figure*}
\centering
\includegraphics[width=\linewidth]{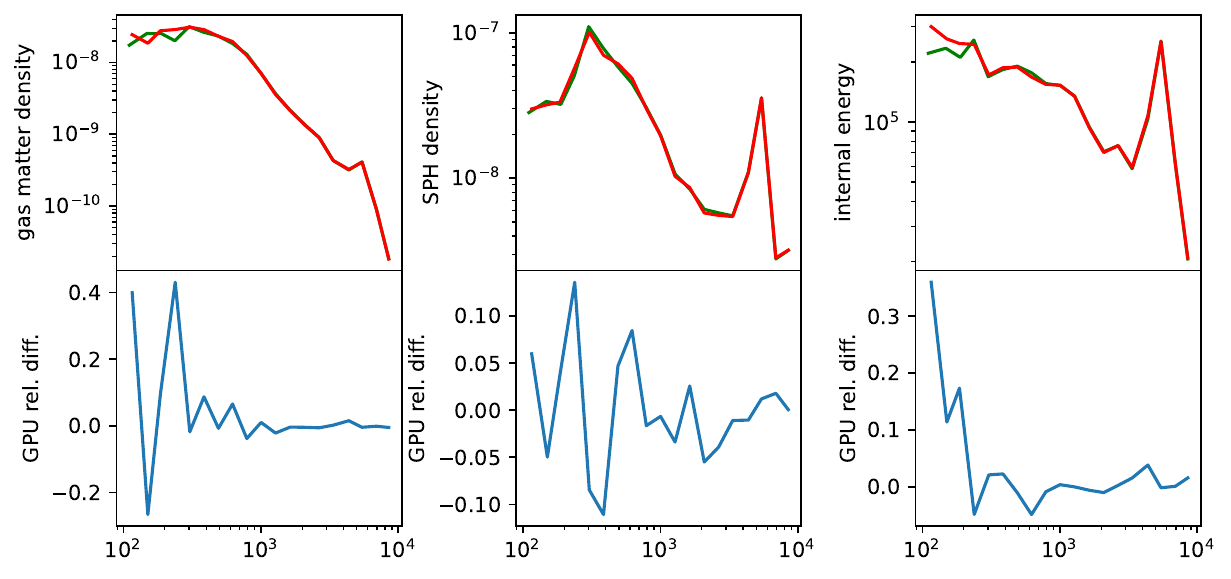}
\caption{\label{fig:norad}Non-radiative galaxy simulation. Left panel: gas matter density; middle panel: SPH density (namely, average SPH $\rho$ value); right panel: temperature. Top columns: profiles; bottom columns: relative difference between CPU and GPU.
}
\end{figure*}

In this subsection, we test the implementation of both hydrodynamics and gravity by re-simulating the so-called cluster\footnote{Its initial conditions are taken from the Dianoga set of initial conditions~\citep{2011MNRAS.418.2234Bonafede}.}
Compared to the previous sections, here we switched on thermal conduction~\citep{Jubelgas2004,Dolag2004,Petkova2009}.

The simulations have a resolution level of $\approx2.2\times10^6$ particles, a mass of $M_{\rm vir}\approx10^{12}\,h^{-1}{\rm M}_\odot,$ a dark matter softening $\epsilon_{\rm DM}=3.75\,h^{-1}\,{\rm ckpc},$ and a stellar softening $\epsilon_\star=1.0\,h^{-1}\,{\rm ckpc}.$

We ran this test on a single node of the LMU CIP cluster.
In both CPU and GPU cases, we used $1$ MPI rank with $12$ OpenMP threads, while the GPU run also utilised one GPU per MPI rank (see the third row in Table~\ref{tab:tests}).

At the end of the simulation, the SPH computation took $8.5\times10^{3}\,{\rm s}$ on the CPU and  $3.2\times10^{3}\,{\rm s}$ on the GPU (speed-up of $2.6$), while the gravitational short-range interaction took $1.5\times10^4\,{\rm s}$ on the CPU and $6.8\times10^3\,{\rm s}$ (speed-up of $22$),  the total time spent by the CPU is $1.6\times10^4\,{\rm s},$ while the GPU spent 
 $1.5\times10^3\,{\rm s}$ (total speed-up of $10$).
Note that these test cases serve to test the accuracy and fit one single GPU, the large speed-up here is due to the lack of communication overhead.

In Figure~\ref{fig:norad_map}, we show the projected density of the resulting cluster (upper panel), together with the bin-wise relative difference PDF (lower panel), where we can see that most of the bins are well below $1\%$ accuracy.
To analyse in more detail, in Figure~\ref{fig:norad}, we show the gas density profile $\rho_{\rm gas}$ (namely the gas mass divided by the shell volume), the mean radial SPH density (namely the average of gas particle densities as stored by \og), and the temperature $T$, where we confirm that the two profiles match for the three quantities between the CPU and the GPU simulation within a few percent at large radii, while they can diverge in the core of the cluster. Note that run-to-run deviations in the cluster core are expected, due to the high density and the amplified impact of small timestep differences accumulated throughout the simulation.

\subsection{Full physics galaxy cluster}
\label{sec:fp}

\begin{figure}
{\centering
\hspace{1cm}CPU\hspace{3cm}GPU\\
\includegraphics[width=\linewidth]{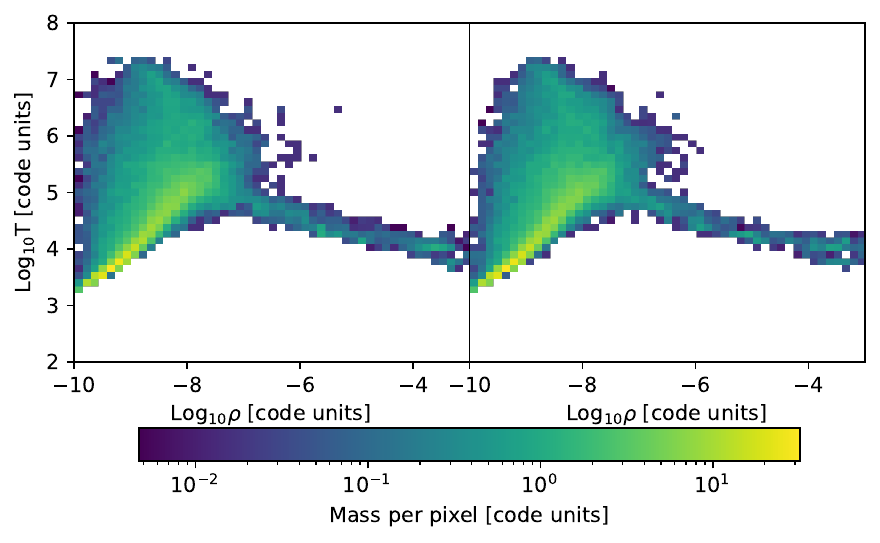} \\
\includegraphics[width=\linewidth]{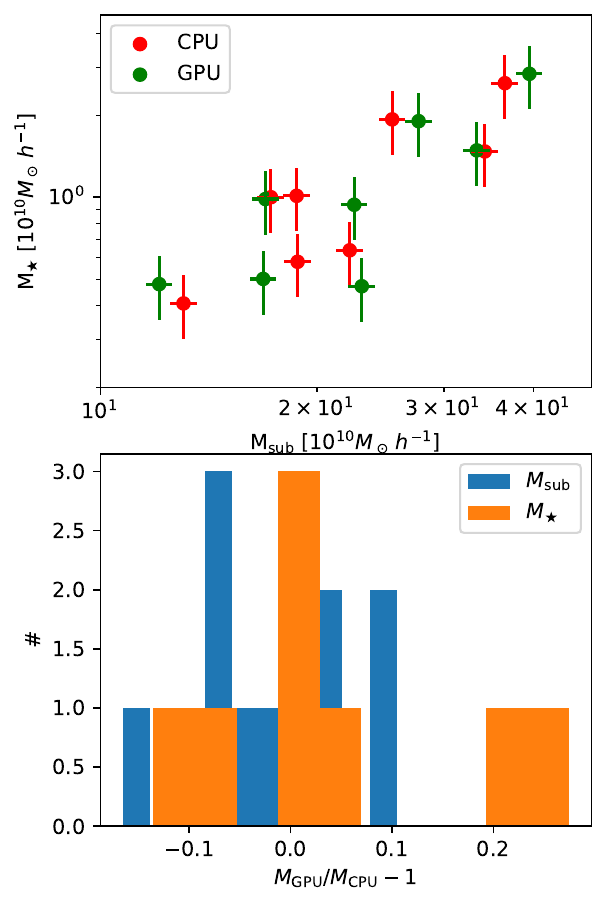}
}
\caption{
CPU vs. GPU runs of the full-physics cluster.
Top panel: the gas phase-space map for the whole high-resolution region of the simulation. The central panel shows the stellar mass vs. halo mass of the formed galaxies, and the bottom panel shows the relative difference between CPU and GPU runs.
}\label{fig:asin}
\end{figure}

We then ran the test from the previous Section with the addition of cooling~\citep{Wiersma2009,Tornatore2010}\footnote{This test additionally used HDF5 for reading cooling tables~\citep{HDF5}}, star formation~\citep{SpringelHernquist2003}, stellar evolution~\citep{Tornatore2004,Tornatore2007}, and black hole physics~\citep{Springel2005Gadget2,SpringelDiMatteo2005,Fabjan2010,Hirschmann2014,Steinborn2015,Damiano2024}.
We used the same MPI and OpenMP setup as in Section~\ref{sec:norad}; namely, we ran this test on a single node of the LMU CIP cluster to exploit one GPU at its best, and we used $1$ MPI rank with $12$ OpenMP threads. The GPU run used one GPU per MPI rank (see fourth row in Table~\ref{tab:tests}).

The SPH module took $1.11\times10^3\,{\rm s}$ in the CPU and  $.56\times10^3\,{\rm s}$ on the GPU (speedup: $1.96$), the short range gravitational interactions took $19\times10^3\,{\rm s}$ in the GPU and $1.4\times10^3\,{\rm s}$ (speedup: $13$), and the CPU time-to-solution is $21\times10^3\,{\rm s}$ and the GPU time to solution is $2.4\times10^3\,{\rm s}$ (total speedup: $6.1$).

In the upper part of Figure \ref{fig:asin}, we show the phase-space diagram of temperature vs. density of the SPH particles, where we can see that the thermodynamic properties of the gas closely match between the two runs, even with full physics treatment.

Finally, we report the stellar mass vs. total mass of the central galaxy and its satellites in Figure~\ref{fig:asin} (bottom panel), where we can see that the high stellar-mass satellites' stellar mass fraction shows only small differences between the CPU and GPU runs.
These differences are of the order of magnitude expected from the intrinsic variability (namely, chaotic behaviour) due to the very sensitive AGN mechanism to the displacement of the black hole and the gas density peak \citep[see e.g.][]{Chaitra2026}.

\section{Performance analysis}
\label{sec:progress}

 \begin{figure}
\centering
\includegraphics[width=\linewidth]{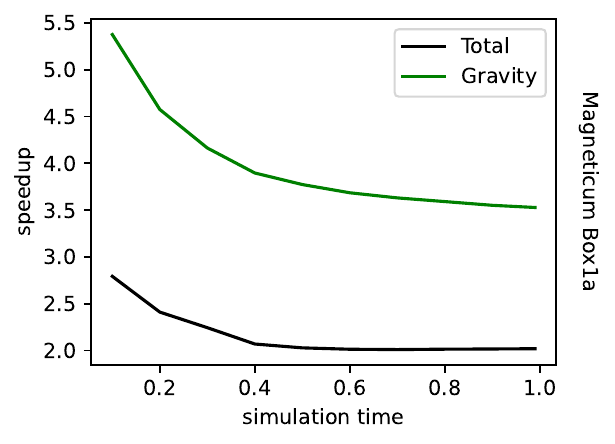}
\includegraphics[width=\linewidth]{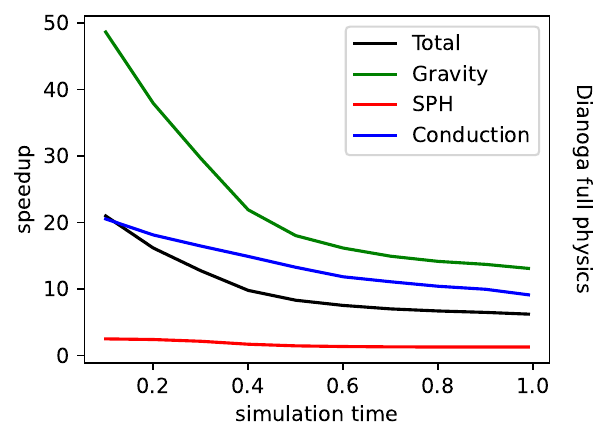}
\caption{\label{fig:time_asin}Time to solution versus scale factor for the Magneticum dark matter-only run (top panel) and the Dianoga asin full-physics run (bottom panel).
}
\end{figure}

In this subsection, we assess the speedup of our GPU modules as the simulations evolve. In particular, we focus on the time spent on Barnes-Hut in the Magneticum DMO run (see Section~\ref{sec:dmo}) and on both Barnes-Hut and SPH in the Dianoga asin/1x run (see Section~\ref{sec:fp}).
The speedups are shown in the top and bottom panels, respectively, of Figure~\ref{fig:time_asin} as the simulation time progresses from 0 to 1.
We can see that the gravitational force achieves the largest speedup; however, because of various overheads in the code, the total speedup stays above $\approx 2.$
In order to maximise the speedup, we suggest filling as many particles as possible on each GPU (as happens in the Dianoga test, bottom panel), which is capable of providing a speedup of $\approx 5.$

\subsection{Scaling}
\label{sec:scaling}

\begin{figure}
\centering
\includegraphics[width=\linewidth]{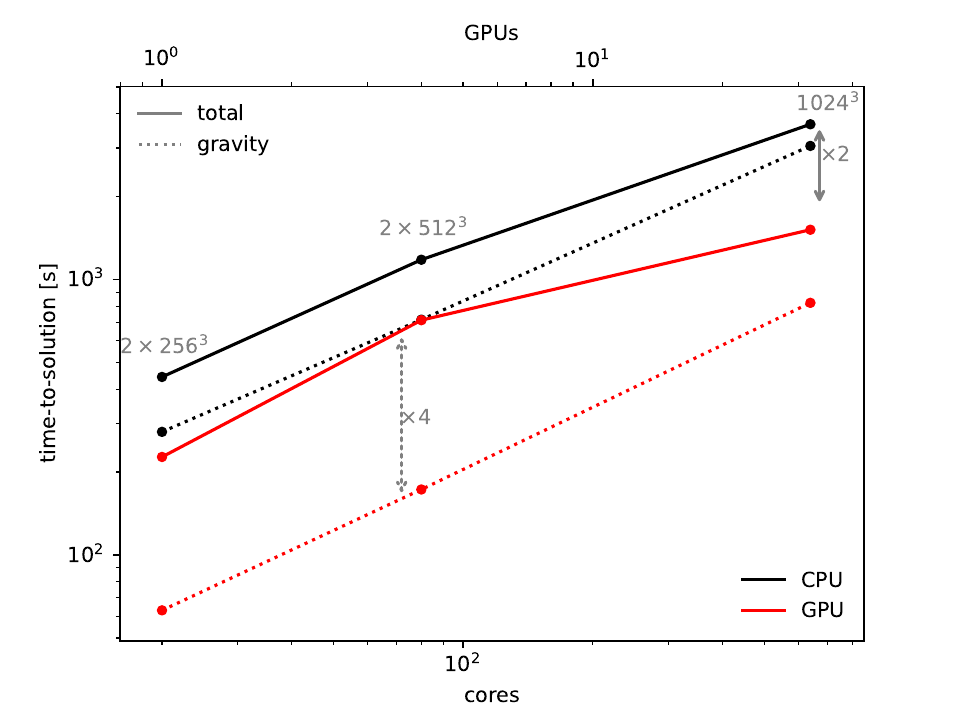}
\caption{\label{fig:scaling}\og chip-to-chip scaling on MareNostrumV-ACC. We used $2\times256^3$, $2\times512^3,$ and $1024^3$ particles, respectively. Both setups used $20$ OpenMP threads per MPI rank. The bottom axis shows the number of cores used in both CPU and GPU runs, while the top axis shows the number of GPUs used in the GPU run. The simulations consist of 100 steps of gravity-only time integration from a very early and homogeneous stage of the simulations. Solid lines show the total times for CPU (black, upper), GPU (red, lower), and the dashed line shows the respective values for the short-range interaction.
}
\end{figure}

At the 7th MareNostrum BSC hackathon\footnote{\url{https://www.bsc.es/news/events/mnhack25-7th-marenostrum-hackathon}}, we performed comparison runs of the OpenACC porting with initial conditions of sizes $2\times256^3$, $2\times512^3$ and $1024^3$ particles (see Fig. \ref{fig:scaling}; note that the last test case was performed without gas particles), where we found a consistent total speedup of 2, while the short-range gravity interaction speedup can reach a factor of $\sim 4$. We caution that this occurs for very homogeneous initial conditions, and the situation may worsen for more clustered conditions.

In addition, large-scale tests have been performed on the Leonardo Booster \citep[][]{SPACELeonardo2025} utilising up to 8192 GPUs, demonstrating consistent scaling behaviour, maintaining over $80\%$ efficiency up to a 32$\times$ resource increase.

\begin{figure}
\centering
\includegraphics[width=\linewidth]{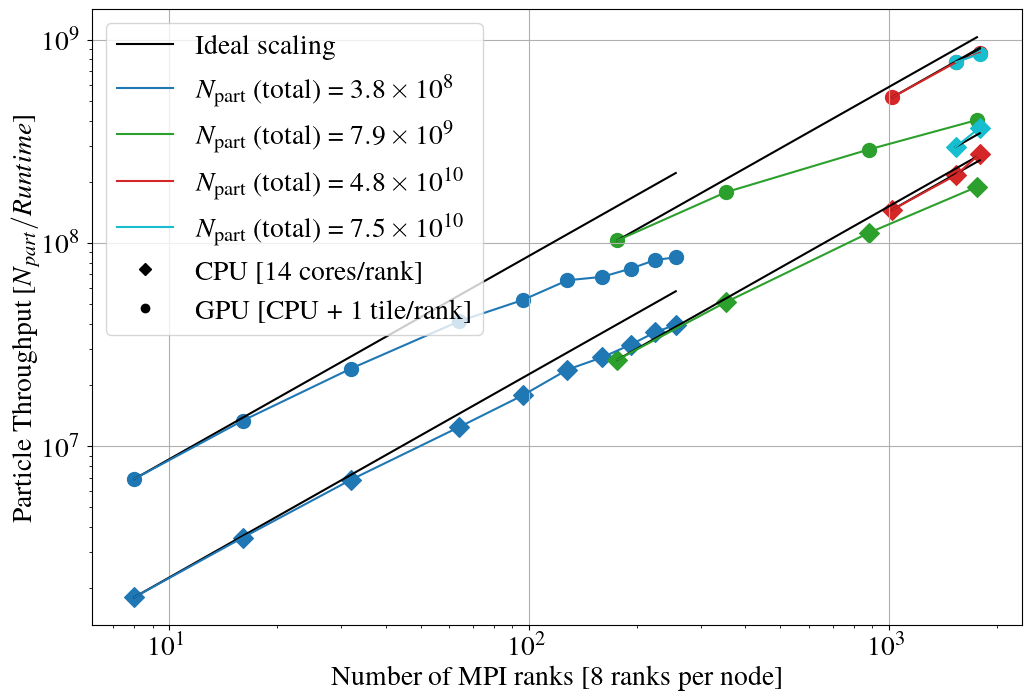}
\includegraphics[width=\linewidth]{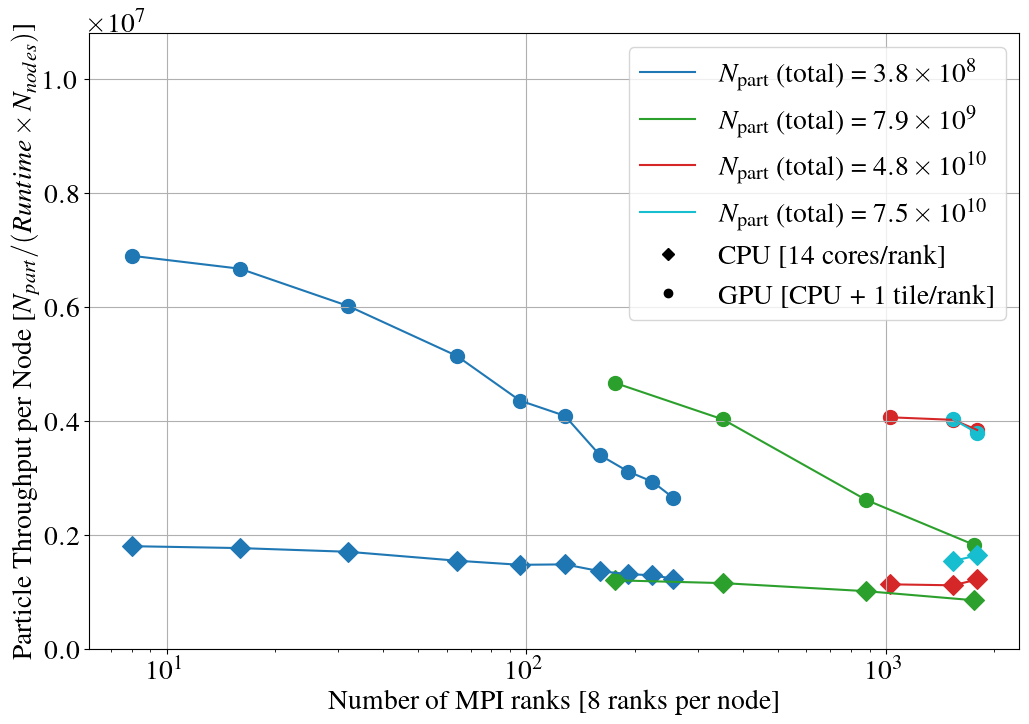}
\caption{Scaling relations for \og on SuperMUC-NG2 using only CPUs or adding GPU accelerators. The top panel shows the strong scaling relation of both implementations, while the lower panel shows the weak scaling results for different test cases.
}
\label{fig:scaling_geray}
\end{figure}

The top panel of Figure~\ref{fig:scaling_geray} shows the strong scaling relation of the OpenMP offloading for the code in comparison with the CPU-only version. While the strong scaling of the CPU-only version is better than that of the GPU version, the code displays an efficiency of $\geq 80\%$ in both cases for an increase in resources by a factor of $\sim8$. In the lower panel, we present the weak scaling results obtained on SuperMUC-NG2 for a dark matter-only run of the Magneticum simulation suite. The GPU and CPU simulations were executed with identical OpenMP/MPI configurations, using one GPU tile per rank in the GPU case. To get a consistent comparison of the different runs, we use the number of particles processed per time and per resource as our metric. The offset between the CPU and GPU runs in the weak scaling plot shows the speedup gained by using the GPUs.

\subsection{Profiling}

\begin{figure*}
\centering
\includegraphics[width=\linewidth]{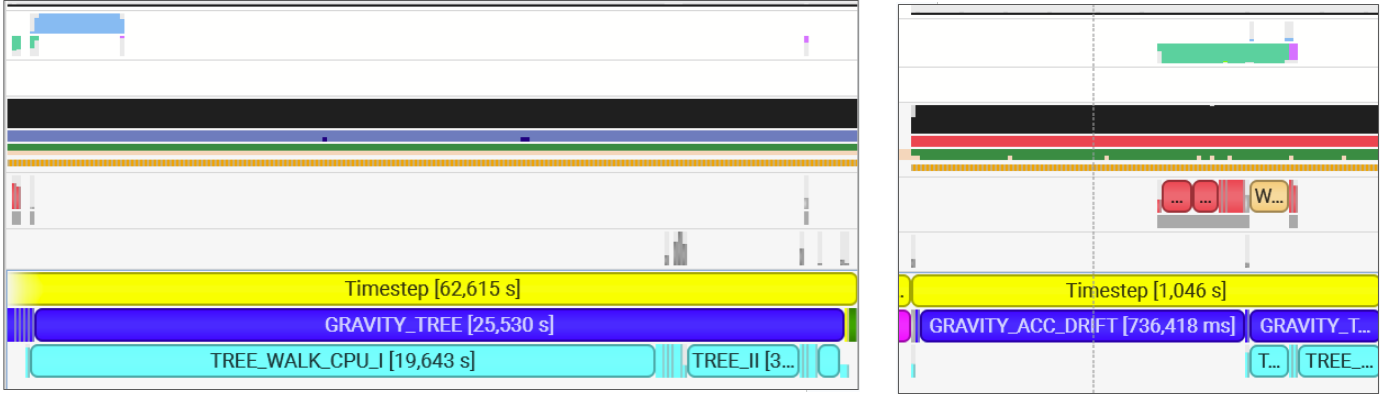}
\caption{\label{fig:nvsight}
Nsight profiling of selected timesteps. Left panel: a large timebin; right panel: a smaller timebin.
The upper part of the profiling shows the CUDA GPU computation timing (light-blue bar) surrounded by upload (green) and download (magenta) of data; the ocher and red boxes in the middle represent similar timings performed by the OpenACC internal timing; the bottom rows represent the internal \og timing performed with~\cite{Karademir2026} library of the full timestep (yellow), the gravity tree interaction module (blue), and the tree walk primary and secondary phases (cyan).
}
\end{figure*}

In Fig.~\ref{fig:nvsight}, we report profiling results using the internal timer hierarchy \citep{Karademir2026} and NVIDIA Nsight Systems tools\footnote{\url{https://developer.nvidia.com/nsight-systems}}, performed on BSC MareNostrum-V. We can see that in the larger timebins of \og, the code is dominated by CPU computation, while on smaller timebins (large enough to be on the GPU) it is dominated by communication. The simulation itself, especially a well-balanced one such as dark matter-only simulations, is dominated by the time of the large timebins, which calls for more optimisations on the CPU side. On the other hand, to improve the overall code performance, in the future we must increase the amount of data on the GPU and minimise host-device transfers.

\section{Conclusions and future prospects}
\label{sec:conclu}

\begin{table*}[b]
    \centering
    \begin{tabular}{r l l l l l}
    \hline
    HPC System & Test Case &   Particles &  Modules tested & Module speedup& Comments\\
    \hline
    SuperMUC-NG & Magneticum Box1a & $2\times1512^3$ & Gravity tree & 4.1\\
        \hline
    LMU CIP & Ryu Jones & $6.3\times10^5$ & SPH & 2.6\\ 
    \hline
    LMU CIP & non-rad Dianoga & $6.3\times10^5$ & Gravity & 10 & 1MPI rank run to stress the GPU\\ 
     &&& SPH & 22& 1MPI rank run to stress the GPU\\ 
    \hline
   LMU CIP & full physics Dianoga & & Gravity & 1.96 \\ 
    & & & SPH+Conduction & 6.1 & \\
    \hline
    \end{tabular}
    \caption{Review of tests and speedup}
    \label{tab:tests}
\end{table*}

In this manuscript, we presented a number of performance and accuracy tests on the GPU porting of \og based on an improved implementation originally presented in~\citep{Ragagnin2020GPU}. In Table \ref{tab:tests}, we review the tests that we presented, together with their node configuration and speedups, where we can see they are always greater than $2$.

The future plans will move in various directions.
One direction is to port new modules to GPUs, for instance, cooling, star formation, and stellar feedback, together with adding support for OpenMP-target, given that OpenMP seems to be the preferred accelerator paradigm on non-NVIDIA GPUs.

Moreover, currently we offload particle data with an array-of-structures (AoS) data layout because we recycled the same code infrastructure that Gadget-like codes use to compute inter-node forces. However, in the future, moving to a structure-of-arrays (SoA) layout, or SoA of small structures (SoAoS) layout, should improve GPU vectorisation and overall performance.

Other benefits could come from additional GPU-wise tuning to exploit modern GPU-based machines even better.
For instance, increasing the amount of data that is persistent in GPU memory, or optimising the tree walk with techniques similar to~\cite{Ragagnin2016GreenTree}.

\section*{Acknowledgements}

We thank the anonymous referees, in particular referee \#1 whose comments contributed in improving the GPU porting.
KD and FG acknowledge support by the COMPLEX project from the European Research Council (ERC) under the European Union’s Horizon 2020 research and innovation program grant agreement ERC-2019-AdG 882679 (COMPLEX) as well as support by the Deutsche Forschungsgemeinschaft (DFG, German Research Foundation) under Germany’s Excellence Strategy - EXC-2094 - 390783311 (ORIGINS). KD also acknowledges support from the European High Performance Computing Joint Undertaking (JU) and Belgium, Czech Republic, France, Germany, Greece, Italy, Norway, and Spain under grant agreement No 101093441 (SPACE). We acknowledge the CINECA award under the ISCRA initiative, for the availability of high-performance computing resources and support (project IsCb1 openaccg). We acknowledge the EuroHPC Joint Undertaking for awarding access to the supercomputer Leonardo BOOSTER, hosted by CINECA (Italy), with the benchmark call project EHPC-BEN-2023B08-013. The authors gratefully acknowledge the scientific support and HPC resources provided by the Erlangen National High Performance Computing Center (NHR@FAU) of the Friedrich-Alexander-Universität Erlangen-Nürnberg (FAU) under the NHR project b123dd. NHR funding is provided by federal and Bavarian state authorities. NHR@FAU hardware is partially funded by the German Research Foundation (DFG) – 440719683. We acknowledge EuroHPC Joint Undertaking for awarding the project ID EHPC-REG-2024R01-029 access to Leonardo at CINECA, Italy.
We acknowledge EuroHPC Joint Undertaking for awarding us access to MareNostrum5 as BSC, Spain (ID EHPC-BEN-2025B11-033) and LUMI at CSC, Finland (ID EHPC-DEV-2026D02-240).
We acknowledge ISCRA for awarding this project access to the LEONARDO supercomputer, owned by the EuroHPC Joint Undertaking, hosted by CINECA, Italy (HP10BUFI59).

\bibliographystyle{plainnat}
\bibliography{sample}

\appendix

\section{Code porting structure}
\label{ap:code}
In this Section, we detail the code changes performed during the porting of \og to GPU for the modules described in this paper (gravity module in \verb|Gravity/gravtree.cpp|, SPH HSML finding in \verb|Hydro/find_hsml.cpp|, SPH gradients computation in \verb|Hydro/gradients.cpp|, SPH hydro-force computation in \verb|Hydro/hydra.cpp|, and thermal conduction in \verb|Hydro/conduction.cpp|). All changes are enveloped within C defines starting with "\verb|ACC_|" so when they are not defined, the code is the same as the standard one; most of the OpenACC-related code is stored in the folder \verb|OpenACC/| for easy readability.

Our drift strategy is implemented in the simulation main loop (in \verb|CodeBase/run.cpp|), in the function that drifts particles (\verb|find_sync_point_and_drift|). If the current time-step has enough active particles (C-defined \verb|ACC_ACTIVEPART_CPU|), then a flag in the OpenACC config variable \verb|acc_all| is raised (\verb|acc_all.go_gpu=1|) and all particles and tree nodes are drifted on a parallel loop, calling \verb|drift_particle(i, All.Ti_Current)| and \verb|force_drift_node(no, All.Ti_Current)| for each particle \verb|i| and node \verb|no|, respectively.

The offload of the tree is performed in the tree-gravity module (in \verb|Gravity/gravtree.cpp|, function \verb|compute_gravity|). There, if \verb|acc_all.go_gpu=1| and the tree has been rebuilt during the current time-step, we explicitly exit and then re-enter the OpenACC data region in order to offload the updated tree data structure to the GPU.

Then, for each of the ported modules, if \verb|acc_all.go_gpu=1|, we re-use the Gadget \verb|particle2in_data| routines (that forge an array of structures of input data for target particles that are exported to other MPI ranks) to prepare an AoS \verb|DataIn| (of the same struct type as the one exported) that will be processed as the Gadget particle data \verb|P| by the functions that evaluate interactions (e.g., \verb|forctree_evaluate_shortrange|, or \verb|density_evaluate|). Similarly, we allocate space for an AoS \verb|DataOut| structure (of the same struct type as the one sent back to the original MPI rank in the CPU-only runs). This allows us to offload to GPU only the minimally required input and output quantities for the active and possible neighbouring particles. Moreover, we can do it very easily by using the already-available functions to fill the MPI communication input and output SoAs.

The module then starts an asynchronous kernel on the GPU, which, in the case of the hydro modules, loops over a chunk of $32$ neighbours (tunable C-defined \verb|ACC_GPU_NGB|). The GPU kernel then calls the same function used on the CPU. This can be done even if the evaluate function sees the \verb|P| variable of a different type on the CPU (as \verb|particle_data_structure|) and on the GPU (as \verb|DataIn| type), thanks to the fact that the same function is compiled on the CPU and the GPU by sharing the same body and having two different signatures. For instance, the hydra evaluate function is defined as follows:

\begin{lstlisting}[basicstyle=\small]
#ifdef ACC_INCLUDE_HYDRA
/* here below, the GPU signature of the
 hydro_evaluate function, which will be compiled
 as hydro_evaluate_local */
#pragma acc routine 
int hydro_evaluate_local(
  int target,
  int p_target,
  int mode,
  int *ngblist,
  int numngb,
  struct hydrodata_in *HydroDataGet,
  struct hydrodata_out *HydroDataResult,
  // we pass a smaller data structre instead of
  // the global variable P
  struct gravity_walkdata *P,
  struct hydrodata_in *SphP,
  ...)
#else
/* here below, the CPU signature of the hydro_evaluate
 function, note that we do not pass P and SphP, since
 in legacy Gadget code, these variables are global */
int hydro_evaluate(
 int target,
 int mode,
 int *ngblist, 
 int extended_numngb_inbox,
 int *i_startnode,
 int node,
 int i_node,
 void *data)                   
#endif 
{
 /* CPU and GPU function body source codes will
 produce P and SphP reference that points either to
 the global data structures (on CPU) or the smaller
 ones passed as the argument (on GPU) */
}
\end{lstlisting}

Meanwhile, the CPU code will run its primary phase function, with an additional parameter \verb|drift_all| that will warn the tree walk that, if this timestep is running on GPU, there is no need to drift any encountered particles (we drifted all particles at the beginning of the timestep). Moreover, since in this scenario the CPU task is  to only fill the communication export buffer, the tree walk is limited to the top nodes by preventing node openening when the top level bitflag of a top node is open, namely \verb|Nodes[no].u.d.bitflags & (1 << BITFLAG_TOPLEVEL)| for a node \verb|no| (see \verb|CodeBase/greentree.hpp| to see how is implemented).
By just adding this filtering line, we leave the overall code almost un-altered, so we could easily extend the existing code base and modules to support this new type of tree walk.

After the module finished the first phase and the secondary phases, the data CPU waits for the GPU to finish (using the \verb|pragma acc wait| directive). The GPU data is then merged to the CPU data using the already existing routine used by the CPU run to merge data received by other MPI ranks, namely the  \verb|out2particle_data| functions).

Another problem that we encountered is that some density module properties are computed as the aggregated minimum of the neighbour ones (e.g. the minimum time bin of the gas particles surrounding a black hole); if a particle has no neighbours in the primary phase, then the secondary phase minimum will be performed against an undefined value.
While in a CPU run, this is virtually impossible, this happens constantly in the GPU-ported code because the primary phase is executed asynchronously, and it is possible that the secondary phase (that has some chances to find zero neighbours for a guest particle) will finish before the GPU computation. Therefore we fixed the code to introduce a new flag for each particle,  to store if the particle had neighbours or not.

\end{document}